\pdfoutput=1
\documentclass[11pt]{article}

\usepackage[]{acl}

\usepackage{times}
\usepackage{latexsym}

\usepackage[T1]{fontenc}
\usepackage[utf8]{inputenc}

\usepackage{microtype}

\usepackage{inconsolata}

\usepackage{graphicx}
\usepackage{times}
\usepackage{latexsym}

\usepackage[normalem]{ulem}
\usepackage{mathrsfs}
\usepackage{amsthm}
\usepackage{amsmath}
\usepackage{amssymb, bm}
\usepackage{multirow}
\usepackage{booktabs}
\usepackage{enumitem}
\usepackage{times}
\usepackage{latexsym}
\usepackage{array}
\usepackage{multirow}
\usepackage{graphicx}
\usepackage{amsfonts}
\usepackage{subcaption}
\usepackage{xspace}
\usepackage{tabularx}
\usepackage{algorithm}
\usepackage{arydshln}
\usepackage{algpseudocode}
\usepackage{colortbl}
\usepackage{tcolorbox}
\usepackage{pifont} 

\definecolor{mypurple}{RGB}{111,61,121}
\definecolor{myblue}{RGB}{46,88,180}
\definecolor{myred}{RGB}{181,68,106}
\definecolor{textorange}{RGB}{237,125,49}
\definecolor{textblue}{RGB}{46,117,181}
\definecolor{textgreen}{RGB}{112,173,71}
\definecolor{myred}{RGB}{181,68,106}

\newcommand{\green}[1]{{\color{textblue}{{#1}}}}
\newcommand{\ours}{ThinkQE\xspace}

\title{ThinkQE: Query Expansion via an Evolving Thinking Process}

\author{Yibin Lei\textsuperscript{1}, Tao Shen\textsuperscript{2}, Andrew Yates\textsuperscript{3} \\
\textsuperscript{1}{University of Amsterdam} \quad \textsuperscript{2}{University of Technology Sydney} \\
\textsuperscript{3}{Johns Hopkins University, HLTCOE} \\
\texttt{y.lei@uva.nl}, \texttt{tao.shen@uts.edu.au},
\texttt{andrew.yates@jhu.edu}
}

\begin{document}
\maketitle
\begin{abstract}
Effective query expansion for web search benefits from promoting both exploration and result diversity to capture multiple interpretations and facets of a query. While recent LLM-based methods have improved retrieval performance and demonstrate strong domain generalization without additional training, they often generate narrowly focused expansions that overlook these desiderata. We propose \ours, a test-time query expansion framework addressing this limitation through two key components: a thinking-based expansion process that encourages deeper and comprehensive semantic exploration, and a corpus-interaction 
strategy that iteratively refines expansions using retrieval feedback from the corpus. Experiments on diverse web search benchmarks (DL19, DL20, and BRIGHT) show \ours consistently outperforms prior approaches, including training-intensive dense retrievers and rerankers.\footnote{Our code is publicly available at \url{https://github.com/Yibin-Lei/Think_QE}.}
\end{abstract}

\section{Introduction}
Query expansion (QE) is a common practice in web search scenarios~\cite{robertson1990termselection, qiu1993conceptbased}, particularly for first-stage retrievers such as BM25~\citep{robertson1995okapi}. Effective expansion involves not only reinforcing the core intent of the query but also introducing terms that capture different facets or interpretations of the information need. 
This multifaceted coverage helps capture a broader semantic context, enabling the retrieval of a more comprehensive set of relevant documents. 
Prior studies have shown that such broad-coverage expansion strategies lead to substantial improvements in retrieval quality~\cite{Bouchoucha2013diversify}.

Recent advances in large language models (LLMs) have led to strong performance in query expansion~\cite{hyde,query2doc,prompt_qe,generative_relevance_feedback,shen2024retrieval,lei-etal-2024-corpus}, particularly due to their ability to rapidly adapt to new domains without requiring additional training. However, existing LLM-based methods often pay limited attention to exploration and result diversity. As illustrated in Table~\ref{tab:case_study}, we observe that current approaches, such as HyDE, tend to generate overly confident expansions that focus narrowly on a single interpretation of the input query. This behavior can be attributed to the model's reliance on its internal knowledge and high-probability completions~\cite{yona2024canlarge,ohi2024likelihood,sun2025largelanguagemodelsoverconfident}, which may suppress alternative formulations or less common aspects of the query. This lack of breadth limits the retrieval of documents reflecting alternative scenarios or requiring more nuanced reasoning.

\begin{table}[t]
\setlength{\belowcaptionskip}{-0.5cm}
        \resizebox{0.95\columnwidth}{!}{
        \begin{tabular}{p{11cm}}
        \toprule
        
        \textbf{\textit{Query}}: Who is robert gray \\
        \midrule
         \textbf{\textit{Expansion w/o. Thinking:}} \\ 
         Robert Gray is best known as the American captain who discovered the Columbia River in 1792. He named the river after his ship, the Columbia Rediviva, and explored it up to Grays Bay. His discovery was later documented by Lieutenant William Broughton during the Vancouver expedition. \\
         \midrule
         \textbf{\textit{\ours:}} \\
         Robert Gray is best known as Captain Robert Gray, \green{an American explorer who played a significant role in the exploration of the Pacific Northwest}. In 1792, he captained the ship Columbia Rediviva and became the first American to navigate the Columbia River, which he named after his vessel. \green{On May 11}, 1792, he entered the mouth of the river and explored approximately \green{20 miles upstream as far as Grays Bay}, which was later named in his honor by Lieutenant William Broughton of the Vancouver expedition. \green{This expedition contributed to the mapping and understanding of the region, highlighting Gray's importance in early American exploration}.\\
    
        \bottomrule
        \vspace{-3cm}
        \end{tabular}
        }
    \caption{\small Examples comparing a standard expansion with \textit{\ours}, our proposed query expansion method with thinking-augmentation. ThinkQE encourages deeper reasoning and multifaceted contextualization. 
    }
    \label{tab:case_study}
    \end{table}

To address these limitations, we propose \ours, a new framework that improves exploration and result diversity along two complementary dimensions. First, we introduce a \textit{thinking-based expansion process}, where the model explicitly accumulates intermediate thoughts and hypotheses before producing final expansions. This encourages the emergence of new and more exploratory terms that can help retrieve documents beyond the initial query scope.
Second, inspired by pseudo-relevance feedback~\cite{rocchio1971relevance}, we propose an \textit{interactive expansion strategy},
where query expansions are progressively refined using feedback from the documents retrieved at each stage. This dynamic interaction with the corpus allows the query to evolve in a context-aware manner, adapting to newly retrieved evidence.

By combining both components, we develop \ours, a test-time query expansion method that achieves strong performance on natural language web search benchmarks, including DL19, DL20, and the StackExchange domain of the BRIGHT benchmark\footnote{We omit math and coding subsets, as ThinkQE relies on natural language expansions, which may not be well-suited for symbolic or structured domains.}.
Remarkably, \ours requires no additional training, yet surpasses recent training-intensive reranking methods, including those based on reinforcement learning and distillation from DeepSeek-R1.  Our analysis reveals that: (1) explicitly modeling a thinking process enhances expansion quality, and (2) iteratively refining queries with evolving retrieval feedback is more effective than generating static expansions, even under the same compute budget.

\section{Method}
\label{sec:method}

We introduce \ours, a query expansion framework that tightly integrates LLM-based thinking process with evolving corpus interaction. \ours follows prior work in generating query expansions using retrieved documents but distinguishes itself through its design of thinking augmentation and iterative corpus interactions.
The method is designed to enable exploration of the query space through thinking processes and evolving refinement based on retrieval feedback from the corpus.
The overall process proceeds in multiple rounds. At each round, an LLM performs thinking-augmented expansion based on the original query and newly retrieved documents from the corpus, which in turn informs subsequent retrieval and expansion steps. 
The following subsections describe each component of the method in detail.

\subsection{Retrieving Initial Evidence from Corpus}
\label{sec:initial_retrieval}

Let $q_0$ denote the original user query. To ground the expansion process in corpus evidence, we begin by retrieving an initial set of documents from the corpus $\mathcal{C}$ using a first-stage lexical retriever. In our implementation, we employ BM25. Specifically, we retrieve the top-$K$ documents:
     $\mathcal{D}_0 = \text{TopK}(\text{BM25}(q_0, \mathcal{C})). $

Here, $\mathcal{D}_0$ denotes the ranked list of top-$K$ documents retrieved for $q_0$, ordered by their BM25 relevance scores. This list serves as the initial feedback signal for expansion, providing retrieval-grounded context to the LLM in the first expansion step.

\subsection{Expansion via Thinking Process}
\label{sec:first_expansion}

To produce an initial expansion, we use an R1-distilled LLM trained to generate a thinking chain before answering. 
Given the original query $q_0$ and top-$K$ retrieved documents $\mathcal{D}_0$, the model follows a two-phase process:

1. \textbf{Thinking Phase:} 
The model reflects on $q_0$ and $\mathcal{D}_0$ to identify latent concepts, resolve ambiguities, and surface alternative interpretations or missing aspects of the information need.
    
2. \textbf{Expansion Phase:} Based on the thinking output, the model generates a query expansion segment $e_1$ that builds upon the original query by introducing additional relevant terms and concepts.

Leveraging the R1-distilled model's natural separation of thought and answer allows us to implement the reasoning-expansion workflow without additional scaffolding or prompt engineering. The prompt shown in Table~\ref{tab:prompt} guides the model to generate expansions by thinking over the input query and the top-retrieved documents.

\begin{table}[t]
\setlength{\abovecaptionskip}{0.2cm}
\setlength{\belowcaptionskip}{0.0cm}
\scriptsize
\centering
\begin{tabularx}{\linewidth}{X}
\toprule
\textbf{\ours Prompt} \\
\midrule
Given a question "\{$q$\}" and its possible answering passages (most of these passages are wrong) enumerated as: \\
1. $\{d_1\}$;
2. $\{d_2\}$;
3. $\{d_3\}$ \ldots\\
please write a correct answering passage. Use your own knowledge, not just the example passages!\\
\bottomrule
\end{tabularx}
\caption{\small Prompt used in \ours for the thinking-based expansion process. $\{\cdot\}$ denotes the placeholder for the corresponding query and top-K documents.}
\label{tab:prompt}
\vspace{-0.3cm}
\end{table}

\subsection{Evolution via Corpus Feedback}
\label{sec:evolution}

We propose to iterate the above thinking-based expansion. At each round $t = 1, \dots, T$, the method performs the following steps:

1. \textbf{Retrieval:} The current query $q_t$ is used to retrieve a ranked list of documents from the corpus: $\mathcal{R}_t = \text{BM25}(q_t, \mathcal{C}).$

2. \textbf{Redundancy Filtering:} To promote diversity and avoid repetition, we exclude documents that (a) appear in the blocklist $\mathcal{B}_t$, or (b) were among the top-$K$ results in the previous round $\mathcal{D}_{t-1}$. We then select the top-$K$ documents from the remaining candidates: $ \mathcal{D}_t^{\text{new}} = \text{TopK}(\mathcal{R}_t \setminus (\mathcal{B}_t \cup \mathcal{D}_{t-1})). $
The blocklist is updated to include all documents that were filtered out in this round.

3. \textbf{Expansion via Thinking:} The LLM is prompted with the original query $q_0$ and the filtered document set $\mathcal{D}_t^{\text{new}}$ to generate the next expansion $e_{t+1}$, using the same two-phase expansion process described in Section~\ref{sec:first_expansion}.

4. \textbf{Query Update:} The query is iteratively updated by concatenating the new expansion: $q_{t+1} = q_t \oplus e_{t+1}$.

This loop can be repeated for any number of rounds $T$, depending on resource constraints or desired depth. 
Notably, as the query grows longer, successive expansions may dilute or replace the original intent. To mitigate this, we follow~\citet{zhang2024exploringbest} and repeat the original query $n$ times in the final reformulation, with 
$n = \frac{\text{len}(\text{expansions})}{\text{len}(q_0) \times \lambda},
\quad \lambda = 3.$
Here, $\text{len}(\text{expansions})$ refers to the total word count of all expansion segments, and $\text{len}(q_0)$ is the word count of the original query.
This repetition reinforces the core semantics of the original query during iterative refinement.

\paragraph{Remark.}
Our method introduces two key innovations: (1) the use of an explicit, LLM-guided thinking process to encourage deeper exploration during expansion, and (2) an evolving loop that dynamically refines the query based on retrieval feedback. 
Within this evolving process, we design two essential components -- \textit{redundancy filtering} and \textit{expansion accumulation} -- both of which play a critical role in the effectiveness of \ours, as demonstrated in our results in Section~\ref{sec:loop_ablation}.

\section{Experiments}

\subsection{Setup}

\paragraph{Datasets.}
We evaluate \ours on two categories of natural language web search datasets: (1) \textbf{Factoid-style retrieval:} TREC DL19~\citep{dl19} and DL20~\citep{dl20}, widely used benchmarks based on the MS MARCO document collections~\citep{msmarco}; and (2) \textbf{Reasoning-oriented datasets:} The StackExchange domain of the BRIGHT benchmark~\citep{su2025bright}, covering seven diverse sub-domains: {Biology (Bio.)}, {Earth Science (Earth.)}, {Economics (Econ.)}, {Psychology (Psy.)}, {Robotics (Rob.)}, {Stack Overflow (Stack.)}, and {Sustainable Living (Sus.)}. On the BRIGHT benchmark, we omit math and coding datasets to focus on only the StackExchange subsets, as ThinkQE relies on language-model-based natural language expansions, which may not be well-suited for symbolic or structured domains such as code or math.

\paragraph{Implementation.}
We use the QWEN-R1-Distill-14B model~\citep{deepseekai2025deepseekr1incentivizingreasoningcapability} to generate thinking-based query expansions, sampling outputs with a temperature of 0.7. The BM25 retrieval is performed using Pyserini~\cite{pyserini} with default hyperparameters.
At each round, \ours uses the top-5 retrieved documents (truncated to 128 tokens for DL benchmarks and 512 tokens for BRIGHT) to prompt the LLM, and samples 2 candidate expansions to enhance diversity. We set the total number of interaction rounds to 3, for a balance between efficiency and effectiveness. Besides the QWEN-R1-Distill-14B model, we also evalute \ours on a variety of reasoning models, including QWEN3-8B, QWEN3-14B~\cite{yang2025qwen3technicalreport}, OpenThinker2-7B~\cite{touvron2023llama2openfoundation}, and Phi-4-Reasoning-14B~\cite{abdin2025phi4reasoningtechnicalreport}.

\begin{table}[t]
\setlength{\abovecaptionskip}{0.2cm}
\setlength{\belowcaptionskip}{-0.0cm}
\setlength{\tabcolsep}{2pt}
\centering
\resizebox{0.49\textwidth}{!}{
\begin{tabular}{lcccccc}
\toprule
 &  \multicolumn{3}{c}{\textbf{DL19}}          & \multicolumn{3}{c}{\textbf{DL20}} \\
\midrule
   & mAP  & ndcg@10 & R@1k & mAP  & ndcg@10 & R@1k \\
\midrule
BM25   & 30.1 & 50.6 & 75.0 & 28.6 & 48.0 & 78.6  \\
\midrule
\multicolumn{7}{l}{\textit{Supervised Fine-Tuned Dense retrievers}} \\
DPR  & 36.5 & 62.2 & 76.9 & 41.8 & \textbf{65.3} & 81.4 \\
ANCE  & 37.1 & 64.5 & 75.5 & 40.8 & 64.6 & 77.6 \\
Contriever$^\text{FT}$ & 41.7 & 62.1 & 83.6 & 43.6 & 63.2 & 85.8 \\
\midrule
\multicolumn{7}{l}{\textit{R1-Distilled Rerankers on BM25 Top-20 Docs}} \\
Rank1-32B  & - & 64.9 & - & - & 61.2 & - \\
Rank-K-32B  & - & 66.2 & - & - & 64.3 & -    \\
\midrule
\multicolumn{7}{l}{\textit{Zero-shot Query expansions with BM25}} \\
HyDE  & 41.8 & 61.3 & 88.0 & 38.2 & 57.9 & 84.4\\
Query2doc & - & 66.2 & - & - & 62.9 & - \\
MILL  & - & 63.8 & 85.9 & - & 61.8 & 85.3 \\
LameR  & 42.8 & 64.9 & 84.2 & - & - & -\\
CSQE & 43.6 & 63.4 & 87.6 & - & - & - \\
\textbf{{ThinkQE}} \textbf{(\textit{ours})}  \\
\quad \textit{w.} R1-14B & \textbf{45.9} & \textbf{68.8} & \textbf{89.3} & \textbf{43.9} & 64.7 & 87.8 \\
\quad \textit{w.} QWEN3-8B & 44.5 & 65.0 & 87.9 & 41.9 & 62.8 & 88.0 \\
\quad \textit{w.} QWEN3-14B & 45.2 & 64.9 & 88.4 & 42.4 & 63.5 & \textbf{88.4} \\
\quad \textit{w.} OpenThinker2-7B & 44.8 & 65.3 & 87.3 & 43.2 & 63.5 & 87.9 \\
\quad \textit{w.} Phi4-Reasoning-14B & 44.0 & 65.0 & 87.1 & 43.0 & 63.9 & 87.2 \\
\bottomrule
\end{tabular}
}
\caption{Results on TREC DL19 and DL20 datasets. In-domain supervised models DPR, ANCE, and Contriever$^\text{FT}$ are trained on the MS-MARCO dataset and listed for reference. \textbf{Bold} indicates the best result across all models.}
\label{tab:dl}
\end{table}

\begin{table*}[t]
\centering
\resizebox{0.8\textwidth}{!}{
    \begin{tabular}{llrrrrrrrr}
    \hline
           & \textbf{Training}   & \textbf{Bio.} & \textbf{Earth.}& \textbf{Econ. }& \textbf{Psy.}& \textbf{Rob.}& \textbf{Stack.}& \textbf{Sus.}& \textbf{Avg.} \\ \hline
    BM25   & Zero-shot   & 18.2 & 27.9 & 16.4 & 13.4 & 10.9 & 16.3 & 16.1  & 17.0  \\
    BM25 + GPT-4o CoT & Zero-shot   & \textbf{53.6} & 53.6 & 24.3 & 38.6 & 18.8 & 22.7 & 25.9 & 33.9  \\
    \hline
    \textit{LLM-based dense retrievers}  \\
    GritLM-7B    & SFT   & 24.8 & 32.3 & 18.9 & 19.8 & 17.1 & 13.6 & 17.8 &  20.6                \\
    GTE-QWEN-7B     & SFT   & 30.6 & 36.4 & 17.8 & 24.6 & 13.2 & 22.2 & 14.8 & 22.8                \\
    ReasonIR-8B & SFT & 26.2 & 31.4 & 23.3 & 30.0 & 18.0 & 23.9 & 20.5 & 24.8 \\
    \hline
    \textit{Rerankers on BM25 Top-100 docs}  \\
    RankGPT4      & Zeroshot   & 33.8 & 34.2 & 16.7 & 27.0 & 22.3 & 27.7 & 11.1 & 24.7                \\
    RankZephyr-7b     & GPT4-distill   & 21.9 & 23.7 & 14.4 & 10.3 & 7.6 & 13.7 & 16.6 & 15.5                \\
    Rank-R1-14B    & GRPO (RL)  &  31.2 & 38.5 & 21.2 & 26.4 & \underline{22.6} & 18.9 & 27.5  & 26.6  \\ 
    \hline
    \textit{Rerankers on BM25+GPT-4o CoT Top-100 docs} \\
    Rank1-14B       & R1-distill & 49.3 & 37.7 & 22.6 & 35.2 & 22.5 & 20.8 & \textbf{33.6} & 31.7  \\
    Rank-K-32B\textsuperscript{*} & R1-distill & 50.8 & 49.4 & 28.2 & \textbf{46.0} & \textbf{27.3} & \textbf{30.5} & \underline{31.9} & \textbf{37.9} \\
    \hline
    \textit{Query expansion with BM25} \\
    HyDE-R1-14B  & Zero-shot  & 33.3 & 44.9 & 21.1 & 29.8 &  16.3 & 24.1 & 21.0 & 27.2           \\
    LameR-R1-14B & Zero-shot  & 35.1 & 46.1 & 23.7 & 31.0 & 17.7 & 26.4  & 25.3 & 29.3         \\ 
    \textbf{\ours} \textbf{(\textit{ours})} \\
    \quad R1-14B & Zero-shot  &  {47.3} & 52.5 & \underline{29.2} & \underline{40.0} & {19.3} & 28.0 & {27.9}  &  34.9 \\
    \quad QWEN3-8B & Zero-shot & 49.8 & \textbf{55.3} & 27.6 & 36.7 & 19.9 & 29.0 & 28.3 &  35.2 \\
    \quad QWEN3-14B & Zero-shot & 51.5 & 53.2 & 27.8 & 37.2 & 22.0 & 16.1 & 27.5 & 33.6 \\
    \quad OpenThinker2-7B & Zero-shot & 50.5 & \underline{54.1} & 25.8 & 36.7 & 18.1 & 28.2 & 28.9 & 34.6 \\
    \quad Phi-4-Reasoning-14B & Zero-shot & \underline{51.8} & 53.5 & \textbf{29.7} & 38.5 & 21.8 & \underline{29.3} & 27.7 & \underline{36.0} \\
    \hline
    \end{tabular}
}
\vspace{-5pt}
\caption{\small Results on the StackExchange domain of the BRIGHT benchmark in terms of nDCG@10. 
The best and the second best results across all models are in \textbf{bold} and \underline{underlined} font, respectively. 
All models are performed on the original query.
BM25+GPT-4o-CoT refers to using BM25 retrieval results on queries rewritten by GPT-4o with chain-of-thought reasoning traces for reranking.
\textsuperscript{*}Rank-K-32B performs computationally expensive listwise reranking over the top-20 documents.}
\vspace{-5pt}
\label{tab:bright}
\end{table*}

\paragraph{Baselines.}
On DL19 and DL20, we compare \ours to recent SOTA zero-shot query expansion methods 
including HyDE~\citep{hyde}, Query2doc~\citep{query2doc}, MILL~\citep{jia2024mill}, LameR~\citep{shen2024retrieval} and CSQE~\cite{lei-etal-2024-corpus}, which use strong LLMs like text-davinci-003-175B~\cite{ouyang2022traininglanguagemodelsfollow}, GPT-3.5-turbo, LLaMA2-13B-Chat and LLaMA2-70B-Chat~\cite{touvron2023llama2openfoundation}.
We also include two recent rerankers distilled from DeepSeek-R1-685B~\cite{deepseekai2025deepseekr1incentivizingreasoningcapability} thinking traces for comparison: Rank1-32B~\cite{weller2025rank1testtimecomputereranking} and Rank-K-32B~\cite{yang2025rankktesttimereasoninglistwise}.  
For reference, we also report results from supervised dense retrievers trained on MS~MARCO: DPR~\citep{dpr}, ANCE~\citep{ance}, and Contriever$^{\text{FT}}$~\citep{contriever}.

On the BRIGHT benchmark, we consider three categories of baselines: (1) \textbf{LLM-based embedding models} such as {GritLM-7B}~\citep{muennighoff2025generative}, {GTE-Qwen-7B}~\citep{li2023generaltextembeddingsmultistage}, and ReasonIR-8B~\cite{shao2025reasonirtrainingretrieversreasoning}, all trained on massive amounts of retrieval data; (2) \textbf{LLM-based rerankers}, including RankGPT4 (zero-shot)~\citep{sun-etal-2023-chatgpt}, RankZephyr-7B (distilled from GPT-4)~\citep{pradeep2023rankzephyreffectiverobustzeroshot}, Rank1-14B (distilled from DeepSeek-R1-685B)~\citep{weller2025rank1testtimecomputereranking}, Rank-R1-14B (trained via reinforcement learning)~\citep{zhuang2025rankr1enhancingreasoningllmbased}, and Rank-K-32B (distilled from DeepSeek-R1-685B)~\cite{yang2025rankktesttimereasoninglistwise}. Rank1-14B, Rank-R1-14B, and Rank-K-32B explicitly incorporate a thinking process during reranking; and (3) \textbf{Query expansion methods} such as {HyDE} and {LameR}, which use the same underlying model as \ours but do not incorporate any explicit thinking process.\footnote{We provide a detailed analysis of the no-thinking setting for fair comparison with \ours in Section~\ref{sec:thinking_process}.} Our method \ours is evaluated in a zero-shot configuration across all datasets.

\subsection{Main Results}
Results are presented in Tables~\ref{tab:dl} and~\ref{tab:bright}. On DL19 and DL20, \ours outperforms almost all other zero-shot query expansion methods across different underlying models, achieving the highest scores across all metrics. Notably, it performs competitively with supervised dense retrievers such as Contriever$^{\text{FT}}$, despite requiring no additional training. Furthermore, uisng the  QWEN-R1-Distill-
14B model, \ours surpasses R1-distilled reranking models such as Rank1-32B and Rank-K-32B -- which also leverage a thinking process and are significantly more computationally expensive.

On the BRIGHT benchmark, \ours remains the strongest among zero-shot query expansion methods, achieving an average nDCG@10 of 36.0 using the Phi-4-Reasoning-14B model. While Rank-K-32B achieves the highest overall score (37.9), it relies on R1 distillation and listwise reranking over GPT-4o-augmented retrieval results, making it significantly more resource-intensive. In contrast, \ours operates in a fully training-free setting and still outperforms several more expensive rerankers, including RankGPT4 (24.7) and Rank1-14B (31.7). Beyond its efficiency, \ours delivers consistently strong performance across domains, ranking first or second in four sub-domains.

\section{Analysis}
In this section, we conduct a detailed analysis of \ours on the StackExchange domain of the BRIGHT benchmark.

\subsection{Impact of the Thinking Process}
\label{sec:thinking_process}
To evaluate the impact of the thinking process, we conduct two ablation studies on \ours: (1) replacing the used model with its base version, QWEN-14B-Base, which has not been trained to produce reasoning traces, and (2) applying the \textit{NoThinking}~\citep{ma2025reasoningmodelseffectivethinking} method, where we prefill the response with a fabricated thinking block (i.e., \textit{<think>Okay, I think I have finished thinking.</think>}) and allow the model to generate the answer directly from that point. As shown in Table~\ref{tab:thinking_process},  \ours with thinking significantly outperforms both variants, underscoring the importance of generating thinking output. We use the \textit{NoThinking} variant as the main baseline.

\begin{table}[t]
\centering
\footnotesize
\setlength{\tabcolsep}{3.5pt}
\setlength{\abovecaptionskip}{0.2cm}
\setlength{\belowcaptionskip}{-0.0cm}
\begin{tabular}{@{} l ccc @{}}
\toprule
Model & BRIGHT Avg.\\
\midrule
QWEN-14B & 27.6\\
QWEN-R1-14B \textit{w/o. thinking} & 29.8  \\
\midrule
QWEN-R1-14B \textit{w. thinking} & 32.5 \\
\bottomrule
\end{tabular}
\caption{ Impact on the thinking process. We compare three configurations: the base model QWEN-14B without any thinking involved, a \textit{NoThinking} variant that bypasses actual thinking by prefilling a dummy thinking trace, and the \ours model with the thinking process enabled.}
\label{tab:thinking_process}
\end{table}

\subsection{Impact of Corpus Interaction}
\begin{figure}[h]
    \centering
    \setlength{\abovecaptionskip}{0.1cm}
    \setlength{\belowcaptionskip}{-0.1cm}
    \includegraphics[width=0.85\linewidth]{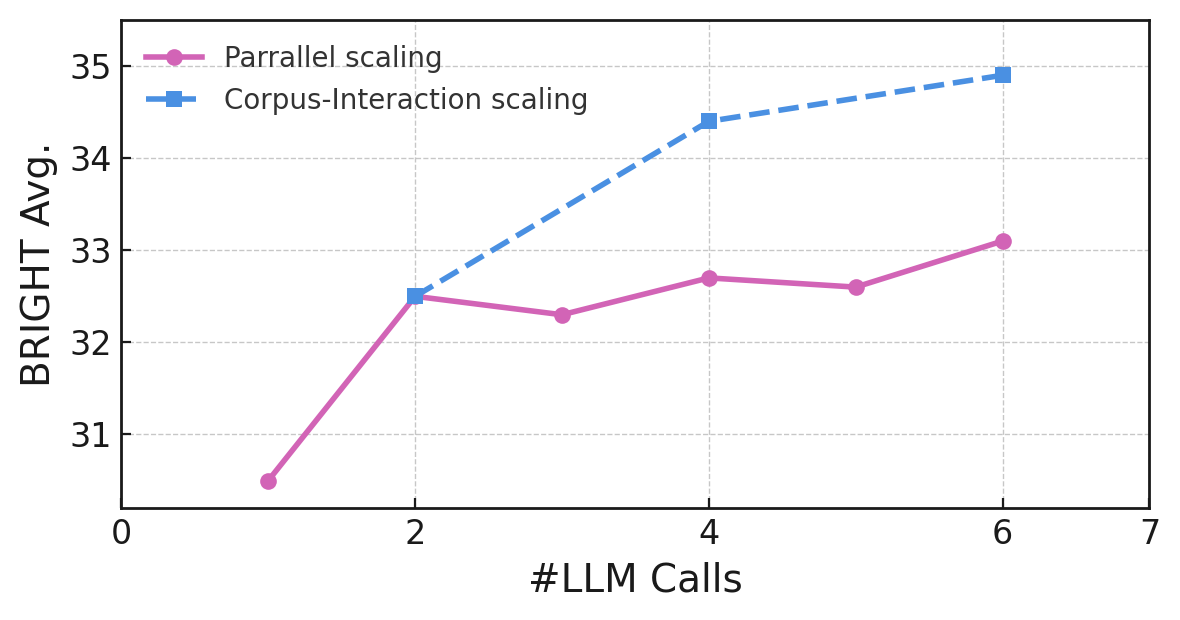}
    \caption{Impact of evolving corpus interaction process.}
    \label{fig:scaling}
    \vspace{-0.0cm}
\end{figure}

To evaluate the corpus interaction process, we compare \ours to a baseline that performs all LLM expansions in a single round -- referred to as parallel scaling. In contrast, \ours uses corpus-interaction scaling, distributing expansions across multiple rounds with retrieval feedback. As shown in Figure~\ref{fig:scaling}, this interaction strategy consistently outperforms the static baseline, indicating that iterative refinement with evolving context is more effective than isolated expansions.

\subsection{Impact on Expansion Accumulation and Redundancy Filter Mechanisms}
\label{sec:loop_ablation}
We conduct a final ablation study on the two core components of the interaction process in \ours: expansion accumulation, where query expansions from different rounds are concatenated to form the new query, and the semantic filter, which excludes top-retrieved documents from the previous round to encourage the introduction of new information. As shown in Table~\ref{tab:abaltion_interaction}, both components are essential for maximizing performance. Disabling either mechanism leads to a noticeable performance drop, highlighting their complementary roles in refining the query and diversifying retrieved evidence across rounds.

\begin{table}[h]
\small
\centering
\setlength{\tabcolsep}{3.5pt}
\setlength{\abovecaptionskip}{0.2cm}
\setlength{\belowcaptionskip}{-0.0cm}
\begin{tabular}{@{} c ccc @{}}
\toprule
Accum. & Filter & BRIGHT Avg.\\
\midrule
\checkmark & \ding{55} & 34.2 \\
\ding{55} & \checkmark & 33.4\\
\midrule
\checkmark & \checkmark & 34.9\\

\bottomrule
\end{tabular}
\caption{ Impact of the expansion accumulation and redundancy filtering mechanisms.}
\label{tab:abaltion_interaction}
\end{table}

\section{Conclusion}
We presented \ours, a test-time query expansion method enhancing exploration and diversity through a thinking-based expansion process and evolving interactions with the corpus. Without requiring any training, \ours consistently improves retrieval performance across multiple benchmarks by encouraging deeper coverage and adaptive refinement, offering a lightweight yet effective alternative to training-based dense retrievers and rerankers.

\section*{Limitations}
The thinking process and evolving interaction process introduce higher inference-time latency and computational cost compared to single-shot expansion methods, which may limit its practicality in latency-sensitive or large-scale deployment scenarios. Furthermore, since our experiments focus exclusively on English web search tasks, the effectiveness of \ours in multilingual settings remains unexplored.

\section*{Acknowledgments}
This research was supported by project VI.Vidi.223.166 of the NWO Talent Programme which is (partly) financed by the Dutch Research Council NWO). 
We acknowledge the Dutch Research Council for awarding this project access to the LUMI supercomputer, owned by the EuroHPC Joint Undertaking, hosted by CSC (Finland) and the LUMI consortium through project number NWO-2024.050.

\bibliography{custom}

@article{robertson1995okapi,
  title={Okapi at TREC-3},
  author={Robertson, Stephen E and Walker, Steve and Jones, Susan and Hancock-Beaulieu, Micheline M and Gatford, Mike and others},
  journal={Nist Special Publication Sp},
  volume={109},
  pages={109},
  year={1995},
  publisher={National Instiute of Standards \& Technology}
}

@article{robertson1990termselection,
author = {Robertson, Stephen},
year = {1990},
month = {12},
pages = {359-364},
title = {On term selection for query expansion},
volume = {46},
journal = {Journal of Documentation},
doi = {10.1108/eb026866}
}

@inproceedings{qiu1993conceptbased,
author = {Qiu, Yonggang and Frei, Hans-Peter},
title = {Concept based query expansion},
year = {1993},
isbn = {0897916050},
publisher = {Association for Computing Machinery},
url = {https://doi.org/10.1145/160688.160713},
doi = {10.1145/160688.160713},
pages = {160–169},
numpages = {10},
address = {Pittsburgh, Pennsylvania},
series = {SIGIR '93},
booktitle={Proceedings of the 16th annual international ACM SIGIR conference on Research and development in information retrieval},
}

@inproceedings{Bouchoucha2013diversify,
author = {Bouchoucha, Arbi and He, Jing and Nie, Jian-Yun},
title = {Diversified query expansion using conceptnet},
year = {2013},
publisher = {Association for Computing Machinery},
url = {https://doi.org/10.1145/2505515.2507881},
booktitle = {Proceedings of the 22nd ACM International Conference on Information \& Knowledge Management},
pages = {1861–1864},
location = {San Francisco, California, USA},
series = {CIKM '13}
}

@article{hyde,
  title={Precise Zero-Shot Dense Retrieval without Relevance Labels},
  author={Gao, Luyu and Ma, Xueguang and Lin, Jimmy and Callan, Jamie},
  journal={arXiv preprint arXiv:2212.10496},
  year={2022},
  url={https://arxiv.org/abs/2212.10496}
}

@article{prompt_qe,
      title={Query Expansion by Prompting Large Language Models}, 
      author={Rolf Jagerman and Honglei Zhuang and Zhen Qin and Xuanhui Wang and Michael Bendersky},
  journal={arXiv preprint arXiv:2305.03653},
  year={2023},
  url={https://arxiv.org/abs/2305.03653}
}

@inproceedings{query2doc,
    title = "Query2doc: Query Expansion with Large Language Models",
    author = "Wang, Liang  and
      Yang, Nan  and
      Wei, Furu",
    booktitle = "Proceedings of the 2023 Conference on Empirical Methods in Natural Language Processing",
    year = "2023",
    address = "Singapore",
    publisher = "Association for Computational Linguistics",
    url = "https://aclanthology.org/2023.emnlp-main.585",
    pages = "9414--9423",
}

@inproceedings{generative_relevance_feedback,
author = {Mackie, Iain and Chatterjee, Shubham and Dalton, Jeffrey},
title = {Generative Relevance Feedback with Large Language Models},
year = {2023},
publisher = {Association for Computing Machinery},
url = {https://doi.org/10.1145/3539618.3591992},
booktitle = {Proceedings of the 46th International ACM SIGIR Conference on Research and Development in Information Retrieval},
pages = {2026–2031},
address = {Taipei, Taiwan},
series = {SIGIR '23}
}

@inproceedings{shen2024retrieval,
    title = "Retrieval-Augmented Retrieval: Large Language Models are Strong Zero-Shot Retriever",
    author = "Shen, Tao  and
      Long, Guodong  and
      Geng, Xiubo  and
      Tao, Chongyang  and
      Lei, Yibin  and
      Zhou, Tianyi  and
      Blumenstein, Michael  and
      Jiang, Daxin",
    booktitle = "Findings of the Association for Computational Linguistics: ACL 2024",
    year = "2024",
    address = "Bangkok, Thailand",
    publisher = "Association for Computational Linguistics",
    url = "https://aclanthology.org/2024.findings-acl.943/",
    pages = "15933--15946",
}

@inproceedings{ohi2024likelihood,
    title = "Likelihood-based Mitigation of Evaluation Bias in Large Language Models",
    author = "Ohi, Masanari  and
      Kaneko, Masahiro  and
      Koike, Ryuto  and
      Loem, Mengsay  and
      Okazaki, Naoaki",
    booktitle = "Findings of the Association for Computational Linguistics: ACL 2024",
    year = "2024",
    address = "Bangkok, Thailand",
    publisher = "Association for Computational Linguistics",
    url = "https://aclanthology.org/2024.findings-acl.193/",
    pages = "3237--3245",
}

@inproceedings{yona2024canlarge,
    title = "Can Large Language Models Faithfully Express Their Intrinsic Uncertainty in Words?",
    author = "Yona, Gal  and
      Aharoni, Roee  and
      Geva, Mor",
    booktitle = "Proceedings of the 2024 Conference on Empirical Methods in Natural Language Processing",
    year = "2024",
    address = "Miami, Florida, USA",
    publisher = "Association for Computational Linguistics",
    url = "https://aclanthology.org/2024.emnlp-main.443/",
    pages = "7752--7764",
}

@article{sun2025largelanguagemodelsoverconfident,
  title={Large Language Models are overconfident and amplify human bias},
  author={Fengfei Sun and Ningke Li and Kailong Wang and Lorenz Goette},
  journal={arXiv preprint arXiv:2505.02151},
  year={2025},
  url={https://arxiv.org/abs/2505.02151}
}

@inproceedings{zhang2024exploringbest,
    title = "Exploring the Best Practices of Query Expansion with Large Language Models",
    author = "Zhang, Le  and
      Wu, Yihong  and
      Yang, Qian  and
      Nie, Jian-Yun",
    booktitle = "Findings of the Association for Computational Linguistics: EMNLP 2024",
    year = "2024",
    address = "Miami, Florida, USA",
    publisher = "Association for Computational Linguistics",
    url = "https://aclanthology.org/2024.findings-emnlp.103/",
    pages = "1872--1883",
}

@inproceedings{pyserini,
author = {Lin, Jimmy and Ma, Xueguang and Lin, Sheng-Chieh and Yang, Jheng-Hong and Pradeep, Ronak and Nogueira, Rodrigo},
title = {Pyserini: A Python Toolkit for Reproducible Information Retrieval Research with Sparse and Dense Representations},
year = {2021},
isbn = {9781450380379},
publisher = {Association for Computing Machinery},
url = {https://doi.org/10.1145/3404835.3463238},
booktitle = {Proceedings of the 44th International ACM SIGIR Conference on Research and Development in Information Retrieval},
pages = {2356–2362},
numpages = {7},
address = {Virtual Event, Canada},
series = {SIGIR '21}
}

@inproceedings{jia2024mill,
    title = "{MILL}: Mutual Verification with Large Language Models for Zero-Shot Query Expansion",
    author = "Jia, Pengyue  and
      Liu, Yiding  and
      Zhao, Xiangyu  and
      Li, Xiaopeng  and
      Hao, Changying  and
      Wang, Shuaiqiang  and
      Yin, Dawei",
    booktitle = "Proceedings of the 2024 Conference of the North American Chapter of the Association for Computational Linguistics: Human Language Technologies (Volume 1: Long Papers)",
    year = "2024",
    address = "Mexico City, Mexico",
    publisher = "Association for Computational Linguistics",
    url = "https://aclanthology.org/2024.naacl-long.138/",
    pages = "2498--2518",
}

@inproceedings{
su2025bright,
title={{BRIGHT}: A Realistic and Challenging Benchmark for Reasoning-Intensive Retrieval},
author={Hongjin Su and Howard Yen and Mengzhou Xia and Weijia Shi and Niklas Muennighoff and Han-yu Wang and Liu Haisu and Quan Shi and Zachary S Siegel and Michael Tang and Ruoxi Sun and Jinsung Yoon and Sercan O Arik and Danqi Chen and Tao Yu},
booktitle={The Thirteenth International Conference on Learning Representations},
year={2025},
url={https://openreview.net/forum?id=ykuc5q381b}
}

@article{dl19,
      title={Overview of the TREC 2019 deep learning track}, 
      author={Nick Craswell and Bhaskar Mitra and Emine Yilmaz and Daniel Campos and Ellen M. Voorhees},
  journal={arXiv preprint arXiv:2003.07820},
  year={2020},
  url={https://arxiv.org/abs/2003.07820}
}

@article{dl20,
      title={Overview of the TREC 2020 deep learning track}, 
      author={Nick Craswell and Bhaskar Mitra and Emine Yilmaz and Daniel Campos and Ellen M. Voorhees},
  journal={arXiv preprint arXiv:2102.07662},
  year={2021},
  url={https://arxiv.org/abs/2102.07662}
}

@article{msmarco,
  url = {https://arxiv.org/abs/1611.09268},
  author = {Bajaj, Payal and Campos, Daniel and Craswell, Nick and Deng, Li and Gao, Jianfeng and others},
  title = {MS MARCO: A Human Generated MAchine Reading COmprehension Dataset},
  journal = {arXiv preprint arXiv:1611.09268},
  year = {2016},
  }

@article{deepseekai2025deepseekr1incentivizingreasoningcapability,
  url = {https://arxiv.org/abs/2501.12948},
  author = {DeepSeek-AI},
  title = {DeepSeek-R1: Incentivizing Reasoning Capability in LLMs via Reinforcement Learning},
  journal = {arXiv preprint arXiv:2501.12948},
  year = {2025},
  }

@article{contriever,
  title={Unsupervised Dense Information Retrieval with Contrastive Learning},
  author={Izacard, Gautier and Caron, Mathilde and Hosseini, Lucas and Riedel, Sebastian and Bojanowski, Piotr and Joulin, Armand and Grave, Edouard},
  journal={Transactions on Machine Learning Research},
  year={2022},
  url={https://openreview.net/forum?id=jKN1pXi7b0}
}

@inproceedings{dpr,
    title = "Dense Passage Retrieval for Open-Domain Question Answering",
    author = "Karpukhin, Vladimir  and
      Oguz, Barlas  and
      Min, Sewon  and
      Lewis, Patrick  and
      Wu, Ledell  and
      Edunov, Sergey  and
      Chen, Danqi  and
      Yih, Wen-tau",
    booktitle = "EMNLP",
    month = nov,
    year = "2020",
    address = "Online",
    url = "https://aclanthology.org/2020.emnlp-main.550",
    doi = "10.18653/v1/2020.emnlp-main.550",
    pages = "6769--6781",
}

@inproceedings{
ance,
title={Approximate Nearest Neighbor Negative Contrastive Learning for Dense Text Retrieval},
author={Lee Xiong and Chenyan Xiong and Ye Li and Kwok-Fung Tang and Jialin Liu and Paul N. Bennett and Junaid Ahmed and Arnold Overwijk},
booktitle={ICLR},
year={2021},
url={https://openreview.net/forum?id=zeFrfgyZln}
}

@inproceedings{
muennighoff2025generative,
title={Generative Representational Instruction Tuning},
author={Niklas Muennighoff and Hongjin SU and Liang Wang and Nan Yang and Furu Wei and Tao Yu and Amanpreet Singh and Douwe Kiela},
booktitle={The Thirteenth International Conference on Learning Representations},
year={2025},
url={https://openreview.net/forum?id=BC4lIvfSzv}
}

@article{li2023generaltextembeddingsmultistage,
  title={Towards General Text Embeddings with Multi-stage Contrastive Learning},
  author={Zehan Li and Xin Zhang and Yanzhao Zhang and Dingkun Long and Pengjun Xie and Meishan Zhang},
  year={2023},
  journal = {arXiv preprint arXiv:2308.03281},
  url={https://arxiv.org/abs/2308.03281}, 
}

@inproceedings{sun-etal-2023-chatgpt,
    title = "Is {C}hat{GPT} Good at Search? Investigating Large Language Models as Re-Ranking Agents",
    author = "Sun, Weiwei  and
      Yan, Lingyong  and
      Ma, Xinyu  and
      Wang, Shuaiqiang  and
      Ren, Pengjie  and
      Chen, Zhumin  and
      Yin, Dawei  and
      Ren, Zhaochun",
    editor = "Bouamor, Houda  and
      Pino, Juan  and
      Bali, Kalika",
    booktitle = "Proceedings of the 2023 Conference on Empirical Methods in Natural Language Processing",
    month = dec,
    year = "2023",
    address = "Singapore",
    publisher = "Association for Computational Linguistics",
    url = "https://aclanthology.org/2023.emnlp-main.923/",
    doi = "10.18653/v1/2023.emnlp-main.923",
    pages = "14918--14937",
}

@article{pradeep2023rankzephyreffectiverobustzeroshot,
      title={RankZephyr: Effective and Robust Zero-Shot Listwise Reranking is a Breeze!}, 
      author={Ronak Pradeep and Sahel Sharifymoghaddam and Jimmy Lin},
      year={2023},
  journal = {arXiv preprint arXiv:2312.02724},
  url={https://arxiv.org/abs/2312.02724}, 
}

@article{weller2025rank1testtimecomputereranking,
      title={Rank1: Test-Time Compute for Reranking in Information Retrieval}, 
      author={Orion Weller and Kathryn Ricci and Eugene Yang and Andrew Yates and Dawn Lawrie and Benjamin Van Durme},
      year={2025},
  journal = {arXiv preprint arXiv:2502.18418},
  url={https://arxiv.org/abs/2502.18418}, 
}

@article{zhuang2025rankr1enhancingreasoningllmbased,
      title={Rank-R1: Enhancing Reasoning in LLM-based Document Rerankers via Reinforcement Learning}, 
      author={Shengyao Zhuang and Xueguang Ma and Bevan Koopman and Jimmy Lin and Guido Zuccon},
      year={2025},
  journal = {arXiv preprint arXiv:2503.06034},
  url={https://arxiv.org/abs/2503.06034}, 
}

@article{ma2025reasoningmodelseffectivethinking,
      title={Reasoning Models Can Be Effective Without Thinking}, 
      author={Wenjie Ma and Jingxuan He and Charlie Snell and Tyler Griggs and Sewon Min and Matei Zaharia},
      year={2025},
  journal = {arXiv preprint arXiv:2504.09858},
  url={https://arxiv.org/abs/2504.09858}, 
}

@article{yang2025rankktesttimereasoninglistwise,
      title={Rank-K: Test-Time Reasoning for Listwise Reranking}, 
      author={Eugene Yang and Andrew Yates and Kathryn Ricci and Orion Weller and Vivek Chari and Benjamin Van Durme and Dawn Lawrie},
      year={2025},
  journal = {arXiv preprint arXiv:2505.14432},
  url={https://arxiv.org/abs/2505.14432}, 
}

@inproceedings{lei-etal-2024-corpus,
    title = "Corpus-Steered Query Expansion with Large Language Models",
    author = "Lei, Yibin  and
      Cao, Yu  and
      Zhou, Tianyi  and
      Shen, Tao  and
      Yates, Andrew",
    booktitle = "Proceedings of the 18th Conference of the European Chapter of the Association for Computational Linguistics (Volume 2: Short Papers)",
    year = "2024",
    publisher = "Association for Computational Linguistics",
    url = "https://aclanthology.org/2024.eacl-short.34/",
    pages = "393--401",
}

@article{ouyang2022traininglanguagemodelsfollow,
      title={Training language models to follow instructions with human feedback}, 
      author={Long Ouyang and Jeff Wu and Xu Jiang and Diogo Almeida and Carroll L. Wainwright and Pamela Mishkin and Chong Zhang and Sandhini Agarwal and Katarina Slama and Alex Ray and John Schulman and Jacob Hilton and Fraser Kelton and Luke Miller and Maddie Simens and Amanda Askell and Peter Welinder and Paul Christiano and Jan Leike and Ryan Lowe},
      year={2022},
  journal = {arXiv preprint arXiv:2203.02155},
  url={https://arxiv.org/abs/2203.02155}, 
}

@article{touvron2023llama2openfoundation,
      title={Llama 2: Open Foundation and Fine-Tuned Chat Models}, 
      author={Meta},
      year={2023},
  journal = {arXiv preprint arXiv:2307.09288},
  url={https://arxiv.org/abs/2307.09288}, 
}

@article{shao2025reasonirtrainingretrieversreasoning,
      title={ReasonIR: Training Retrievers for Reasoning Tasks}, 
      author={Rulin Shao and Rui Qiao and Varsha Kishore and Niklas Muennighoff and Xi Victoria Lin and Daniela Rus and Bryan Kian Hsiang Low and Sewon Min and Wen-tau Yih and Pang Wei Koh and Luke Zettlemoyer},
      year={2025},
  journal = {arXiv preprint arXiv:2504.20595},
  url={https://arxiv.org/abs/2504.20595}, 
}

@article{rocchio1971relevance,
  title={Relevance feedback in information retrieval},
  author={Rocchio Jr, Joseph John},
  journal={The SMART retrieval system: experiments in automatic document processing},
  year={1971},
  publisher={Englewood Cliffs}
}

@article{yang2025qwen3technicalreport,
      title={Qwen3 Technical Report}, 
      author={An Yang and Anfeng Li and Baosong Yang and Beichen Zhang and Binyuan Hui and Bo Zheng and Bowen Yu and Chang Gao and Chengen Huang and Chenxu Lv and Chujie Zheng and Dayiheng Liu and Fan Zhou and Fei Huang and Feng Hu and Hao Ge and Haoran Wei and Huan Lin and Jialong Tang and Jian Yang and Jianhong Tu and Jianwei Zhang and Jianxin Yang and Jiaxi Yang and Jing Zhou and Jingren Zhou and Junyang Lin and Kai Dang and Keqin Bao and Kexin Yang and Le Yu and Lianghao Deng and Mei Li and Mingfeng Xue and Mingze Li and Pei Zhang and Peng Wang and Qin Zhu and Rui Men and Ruize Gao and Shixuan Liu and Shuang Luo and Tianhao Li and Tianyi Tang and Wenbiao Yin and Xingzhang Ren and Xinyu Wang and Xinyu Zhang and Xuancheng Ren and Yang Fan and Yang Su and Yichang Zhang and Yinger Zhang and Yu Wan and Yuqiong Liu and Zekun Wang and Zeyu Cui and Zhenru Zhang and Zhipeng Zhou and Zihan Qiu},
      year={2025},
      journal = {arXiv preprint arXiv:2505.09388},
      url={https://arxiv.org/abs/2505.09388}, 
}

@article{abdin2025phi4reasoningtechnicalreport,
      title={Phi-4-reasoning Technical Report}, 
      author={Marah Abdin and Sahaj Agarwal and Ahmed Awadallah and Vidhisha Balachandran and Harkirat Behl and Lingjiao Chen and Gustavo de Rosa and Suriya Gunasekar and Mojan Javaheripi and Neel Joshi and Piero Kauffmann and Yash Lara and Caio César Teodoro Mendes and Arindam Mitra and Besmira Nushi and Dimitris Papailiopoulos and Olli Saarikivi and Shital Shah and Vaishnavi Shrivastava and Vibhav Vineet and Yue Wu and Safoora Yousefi and Guoqing Zheng},
      year={2025},
      journal = {arXiv preprint arXiv:2504.21318},
      url={https://arxiv.org/abs/2504.21318}, 
}

\newpage
\appendix

\section{Appendix}
\label{sec:appendix}

\subsection{Dataset Statistics}
Details about the retrieval datasets are shown in Table~\ref{tab:data_stat_retrievaaaal}.

\begin{table}[h]
\setlength{\abovecaptionskip}{0.1cm}
\setlength{\belowcaptionskip}{-0.2cm}
\centering
\small
\setlength{\tabcolsep}{3pt}
\begin{tabular}{@{} l rr @{}}
\toprule
Dataset & \#Test & \#Corpus \\
\midrule
DL19 & 43 & 8,841,823 \\
DL20 & 50 & 8,841,823 \\
Biology & 103 & 57,359 \\
Earth Science & 116 & 121,249 \\
Economics  & 103 & 50,220 \\
Psychology    & 101 & 52,835 \\
Robotics    & 101 & 61,961 \\
Stack Overflow & 117 & 107,081 \\
Sustainable Living & 108 & 60,792 \\
\bottomrule
\end{tabular}
\caption{Dataset Statistics.}
\label{tab:data_stat_retrievaaaal}
\end{table}

\subsection{Impact of the Thinking Process}
Results across all domains on the impact of the thinking process are provided in Table~\ref{tab:thinking_details}.

\subsection{Impact of Evolving Corpus Interaction}
Results across all domains on the impact of the evolving corpus interaction are provided in Table~\ref{tab:details_interaction}.

\subsection{Core Components of the Evolving Interaction Process}
Results across all domains on the impact of the expansion accumulation and redundancy filter mechanisms are provided in Table~\ref{tab:details_core_interaction}.

\subsection{Impact of $\lambda$}
Table~\ref{tab:lambda_results} presents detailed results analyzing the impact of $\lambda$, which influences the repetition frequency of the original query during expansion. The results demonstrate that performance differences are small when varying $\lambda$ from 3 to 6. However, lower $\lambda$ values tend to cause excessive repetition of the original query, which generally hurts performance.

\subsection{Effectiveness-Efficiency Analysis on ThinkQE}
We provide latency and performance trade-off results comparing model size scaling, the thinking process, and multiple rounds in Table~\ref{tab:efficiency_results}, using the DeepSeek-R1-Distill-Qwen-14B and DeepSeek-R1-Distill-Qwen-32B models evaluated on a single H100 GPU. Our results show that involving the thinking process and multiple rounds increases latency. However, scaling model size alone yields limited improvements relative to latency increase: moving from R1-14B without thinking (3.71 s/query, 29.8) to R1-32B without thinking (7.88 s/query) improves Bright Avg. by just +0.6 points (30.4) while more than doubling latency. Adding the thinking process, although it increases latency more, is substantially more effective: applying 1 round of thinking to R1-14B boosts performance by +2.7 points (32.5) compared to R1-14B without thinking, with latency rising to 15.40 s/query. Even with thinking enabled, scaling from R1-14B to R1-32B brings only a small additional gain (+0.4 points). Meanwhile, multi-round corpus interaction offers a more efficient path to higher effectiveness, with R1-14B 3-round reaching 34.9 (+2.4 over its 1-round version) and outperforming R1-32B 1-round.

\subsection{ThinkQE on Non-Web Search Datasets}
We evaluate ThinkQE on two additional non-web search datasets, TREC-Covid and Scifact. The results in Table~\ref{tab:beir_results} show that ThinkQE remains highly competitive and often outperforms the baselines. Note that both MILL and LameR are based on the powerful, closed-source GPT-3.5-Turbo model.

\begin{table}[h]
\setlength{\abovecaptionskip}{0.1cm}
\setlength{\belowcaptionskip}{-0.2cm}
\centering
\small
\setlength{\tabcolsep}{3pt}
\begin{tabular}{@{} l rr @{}}
\toprule
Method & TREC-Covid & Scifact \\
\midrule
BM25 & 59.5 & 67.9 \\
Contriever (FT) & 59.6 & 67.7 \\
HyDE & 59.3 & 69.1 \\
Query2Doc & 72.2 & 68.6 \\
MILL & 75.3 & 71.4 \\ 
LameR & 75.8 & 73.5 \\
CSQE &74.2 & 69.6 \\
ThinkQE & 76.1 & 73.3 \\
\bottomrule
\end{tabular}
\caption{Results (NDCG@10) on non-web search datasets.}
\label{tab:beir_results}
\end{table}

\subsection{Significance Test Results of ThinkQE}
We conduct significance testing by comparing ThinkQE with the two most relevant baselines we reimplemented on BRIGHT: HyDE and LameR using the same QWEN-R1-Distill-14B model. The results are presented in Table~\ref{tab:significance_tests}. Significance tests were performed using a t-test with a p-value threshold of 0.05. The results show that ThinkQE is significantly better than HyDE and LameR on 6 out of 7 and 5 out of 7 domains, respectively, demonstrating its effectiveness.

\subsection{Impact of Number of Rounds on Domain-Specific Dataset}

To further analyze the potential topic drift issue during the iterative process, we examine performance changes across one to three rounds using the domain-specific TREC-Covid dataset.  The results, presented in Table~\ref{tab:trec_covid}, demonstrate that increasing from one to two rounds improves the performance, while extending to three rounds largely maintains performance.

\begin{table}[h]
\setlength{\abovecaptionskip}{0.1cm}
\setlength{\belowcaptionskip}{-0.2cm}
\centering
\setlength{\tabcolsep}{3pt}
\begin{tabular}{@{} c r @{}}
\toprule
Round & NDCG@10 \\
\midrule
1 & 75.2 \\
2 & 76.2 \\
3 & 76.1 \\
\bottomrule
\end{tabular}
\caption{Performance across iterative rounds on the domain-specific TREC-COVID dataset.}
\label{tab:trec_covid}
\end{table}

\begin{table*}[h]
\centering
\resizebox{0.8\textwidth}{!}{
    \begin{tabular}{lrrrrrrrr}
    \hline
     & \textbf{Bio.} & \textbf{Earth.}& \textbf{Econ. }& \textbf{Psy.}& \textbf{Rob.}& \textbf{Stack.}& \textbf{Sus.}& \textbf{Avg.} \\ \hline
    QWEN-BASE-14B & 36.7 & 45.1 & 21.9  & 27.7 & 16.8 & 23.3 & 21.7 & 27.6 \\
    QWEN-R1-14B \textit{w/o. thinking} & 39.1 & 45.6 & 25.0  & 30.0 & 18.0 & 26.5 & 24.4 & 29.8 \\
    \midrule
    QWEN-R1-14B \textit{w. thinking} & 42.6 & 50.6 & 26.2  & 35.8 & 18.8 & 28.4 & 25.1 & 32.5 \\
    \hline
    \end{tabular}
}
\caption{Detailed results on the impact of the thinking process.}
\label{tab:thinking_details}
\end{table*}

\begin{table*}[t]
\centering
\resizebox{0.8\textwidth}{!}{
    \begin{tabular}{lrrrrrrrr}
    \hline
     \textbf{\#LLM calls}   & \textbf{Bio.} & \textbf{Earth.}& \textbf{Econ. }& \textbf{Psy.}& \textbf{Rob.}& \textbf{Stack.}& \textbf{Sus.}& \textbf{Avg.} \\ \hline
    \textit{Parallel scaling}  \\
    1&42.6&47.3&25.1&30.3&18.1&24.8&25.2 & 30.5\\
    2&42.6&50.6&26.2&35.8&18.8&28.4&25.1 & 32.5\\
    3&44.2&50.4&26.6&33.6&18.0&26.5&26.5 & 32.3\\
    4&42.4&49.8&27.7&35.5&17.8&28.0&27.4 & 32.7\\ 
    5&41.7&50.7&26.7&35.2&19.3&27.5&27.4 & 32.6\\
    6&45.3&50.3&26.4&34.5&19.0&28.2&28.0 & 33.1\\
    \hline
    \textit{Corpus-interaction scaling}  \\
    2&42.6&50.6&26.2&35.8&18.8&28.4&25.1 & 32.5\\
    4&45.9&52.6&28.3&39.0&18.7&28.5&28.0 & 34.4\\
    6&47.3&52.5&29.2&40.0&19.3&28.0&27.9 & 34.9\\
    \hline
    \end{tabular}
}
\vspace{-5pt}
\caption{Detailed results on the impact of the evolving corpus interaction.}
\vspace{-5pt}
\label{tab:details_interaction}
\end{table*}

\begin{table*}[t]
\centering
\resizebox{0.8\textwidth}{!}{
    \begin{tabular}{llrrrrrrrr}
    \hline
     \textbf{Accum.} & \textbf{Filter}    & \textbf{Bio.} & \textbf{Earth.}& \textbf{Econ. }& \textbf{Psy.}& \textbf{Rob.}& \textbf{Stack.}& \textbf{Sus.}& \textbf{Avg.} \\ \hline
    \checkmark & \ding{55} & 46.4&51.5&27.8&39.5&17.9&28.2&28.0&34.2\\
    \ding{55} & \checkmark & 47.5&50.7&27.9&34.8&17.7&26.5&28.4&33.4\\
    \checkmark & \checkmark & 47.3&52.5&29.2&40.0&19.3&28.0&27.9&34.9 \\
    \hline
    \end{tabular}
}
\vspace{-5pt}
\caption{Detailed results on the impact of the expansion accumulation and redundancy filter mechanism.}
\vspace{-5pt}
\label{tab:details_core_interaction}
\end{table*}

\begin{table*}[h]
\centering
\resizebox{0.6\textwidth}{!}{
    \begin{tabular}{lrrrrrrrr}
    \hline
     $\mathbf{\lambda}$ & \textbf{Bio.} & \textbf{Earth.}& \textbf{Econ. }& \textbf{Psy.}& \textbf{Rob.}& \textbf{Stack.}& \textbf{Sus.}& \textbf{Avg.} \\ \hline
    1 & 38.2 & 47.5 & 25.0 & 32.1 & 18.4 & 25.5 & 24.0 & 30.1 \\
    2 & 44.9 & 50.1 & 28.2 & 37.0 & 19.3 & 27.0 & 26.8 & 33.3 \\
    3 & 46.7 & 51.2 & 29.4 & 39.3 & 20.8 & 28.3 & 28.4 & 34.9 \\
    4 & 48.9 & 52.3 & 28.2 & 38.4 & 20.2 & 28.1 & 29.4 & 35.1 \\
    5 & 49.7 & 52.5 & 29.5 & 37.7 & 19.6 & 28.9 & 27.6 & 35.1 \\
    6 & 49.1 & 51.8 & 29.5 & 39.1 & 19.4 & 29.0 & 28.8 & 35.2\\
    \hline
    \end{tabular}
}
\caption{\small Detailed results on the impact of $\lambda$.}
\label{tab:lambda_results}
\end{table*}

\begin{table*}[h]
\centering
\resizebox{0.6\textwidth}{!}{
    \begin{tabular}{lrrrr}
    \hline
     \textbf{Model} & \textbf{Thinking} & \textbf{Round} & \textbf{Latency (second/query)} & \textbf{Bright Avg.}  \\ \hline
    R1-14B & No & 1 & 3.71 & 29.8 \\
    R1-32B & No & 1 & 7.88 & 30.4 \\
    R1-14B & Yes & 1 & 15.40 & 32.5 \\
    R1-32B & Yes & 1 & 30.53 & 32.9 \\
    R1-14B & Yes & 3 & 45.44 & 34.9 \\
    \hline
    \end{tabular}
}
\caption{Effectiveness-Efficiency Analysis on ThinkQE.}
\label{tab:efficiency_results}
\end{table*}

\begin{table*}[h]
\centering
\resizebox{0.8\textwidth}{!}{
    \begin{tabular}{lcccccccc}
    \hline
     \textbf{Method} & \textbf{Bio.} & \textbf{Earth.}& \textbf{Econ. }& \textbf{Psy.}& \textbf{Rob.}& \textbf{Stack.}& \textbf{Sus.}& \textbf{Avg.} \\ \hline
    HyDE & 33.3 & 44.9 & 21.1 & 29.8 & 16.3 & 24.1 & 21.0 & 27.2 \\
    LameR & 35.1 & 46.1 & 23.7 & 31.0 & 17.7 & 26.4 & 25.3 & 29.3 \\
    \midrule
    ThinkQE & \ \ $47.3^{\dagger\ddagger}$ & \ \ $52.5^{\dagger\ddagger}$ & \ \ $29.2^{\dagger\ddagger}$ & \ \ $40.0^{\dagger\ddagger}$ & 19.3 & \ $28.0^{\dagger}$ & \ \ $27.9^{\dagger\ddagger}$ & 34.9 \\
    \hline
    \end{tabular}
}
\caption{Significance testing results. $\mathbf{\dagger}$ and $\mathbf{\ddagger}$ mean ThinkQE performs significantly better than HyDE and LameR, respectively, as determined by a t-test with p-value 0.05 as threshold.}
\label{tab:significance_tests}
\end{table*}

\end{document}